\documentclass[conference,10pt]{IEEEtran}

\usepackage{caption}
\usepackage{color}
\usepackage{comment}
\usepackage{graphicx}
\usepackage{tikz}
\usepackage{xspace}

\newcommand{\I}[1]{\textit{#1}}
\newcommand{\B}[1]{\textbf{#1}}
\newcommand{\T}[1]{\texttt{#1}}

\newcommand{\paraheader}[1]{\B{#1:}} 

\newcommand{\FIG}[1]{Figure~\ref{figure:#1}}
\newcommand{\CODE}[1]{Figure~\ref{code:#1}}
\newcommand{\PLOT}[1]{Figure~\ref{plot:#1}}

\newcommand{\SL}[1]{~\S\ref{section:#1}}

\newcommand*\circled[1]{\tikz[baseline=(char.base)]{
  \node[shape=circle,draw,inner sep=2pt] (char) {#1};}}

\definecolor{swiftbuiltincolor}{rgb}{0,0,0}
\definecolor{swiftstringcolor}{rgb}{0,0,0}
\definecolor{swiftcommentcolor}{rgb}{0,0,0}

\definecolor{darkblue}{rgb}{0,0,0.7}
\definecolor{darkgreen}{rgb}{0,0.5,0}
\definecolor{orange}{rgb}{0.7,0.5,0.0}
\definecolor{brown}{rgb}{0.8,0.4,0.0}

\newif\ifdraft
\drafttrue
\ifdraft
  \newcommand{\woz}[1]{ {\textcolor{darkgreen} { Wozniak: #1 }}}
  \newcommand{\arm}[1]{ {\textcolor{darkred} { Tim: #1 }}}
  \newcommand{\hemant}[1]{ {\textcolor{darkblue} { Hemant: #1 }}}
  \newcommand{\mw}[1]{ {\textcolor{blue} { Mike: #1 }}}
  \newcommand{\ian}[1]{{\textcolor{red}{ Ian: #1}}}
\else
  \newcommand{\woz}[1]{}
  \newcommand{\arm}[1]{}
  \newcommand{\hemant}[1]{}
  \newcommand{\mw}[1]{}
  \newcommand{\ian}[1]{}
\fi

\begin{document}

\title{Big Data Staging with MPI-IO \\ for Interactive X-ray Science}


\author{\IEEEauthorblockN{
Justin M. Wozniak,\IEEEauthorrefmark{1}\IEEEauthorrefmark{2}
Hemant Sharma,\IEEEauthorrefmark{3}
Timothy G. Armstrong,\IEEEauthorrefmark{4}
Michael Wilde,\IEEEauthorrefmark{1}\IEEEauthorrefmark{2}
Jonathan D. Almer,\IEEEauthorrefmark{3}
Ian Foster\IEEEauthorrefmark{1}\IEEEauthorrefmark{2}\IEEEauthorrefmark{4}
  \IEEEauthorblockA{
  \IEEEauthorblockA{
  \IEEEauthorrefmark{1}Mathematics and Computer Science Division,
    Argonne National Laboratory,
    Argonne, IL, USA}
  \IEEEauthorblockA{\IEEEauthorrefmark{2}Computation Institute,
    University of Chicago and Argonne National Laboratory,
    Chicago, IL, USA}
  \IEEEauthorrefmark{3}X-ray Science Division,
    Argonne National Laboratory,
    Argonne, IL, USA}
  \IEEEauthorrefmark{4}Dept. of Computer Science,
    University of Chicago,
    Chicago, IL, USA}
}

\maketitle

\thispagestyle{plain}
\pagestyle{plain}

\begin{abstract}

New techniques in X-ray scattering science experiments produce large
data sets that can require millions of high-performance processing
hours per week of computation for analysis.  In such applications,
data is typically moved from X-ray detectors to a large parallel file
system shared by all nodes of a petascale supercomputer and then is read
repeatedly as different science application tasks proceed. However,
this straightforward implementation causes significant contention in
the file system.  We propose an alternative approach in which data is
instead staged into and cached in compute node memory for extended
periods, during which time various processing tasks may efficiently
access it.  We describe here such a big data staging framework, based
on MPI-IO and the Swift parallel scripting language.  We discuss a
range of large-scale data management issues involved in X-ray
scattering science and measure the performance benefits of the new
staging framework for high-energy diffraction microscopy, an important
emerging application in data-intensive X-ray scattering.  We show that
our framework accelerates scientific processing
turnaround from three months to under 10 minutes, and that our I/O
technique reduces input overheads by a factor of 5 on 8K
Blue Gene/Q nodes.

\end{abstract}

\section{Introduction}

Many branches of science face a big data challenge as experimental
sensors and simulators produce ever larger data sets.  X-ray
scattering experiments at facilities such as the Advanced Photon Source
(APS) at Argonne National Laboratory (Argonne) are no exception.
Improvements in detector technology can produce $\sim$15~TB per week
or more of raw image data; subsequent processing can more than double that
quantity.  These increases in data sizes outpace increases in
computer performance, creating a crisis situation.  These datasets
will either be used to advance X-ray science in a transformative way,
or be discarded.

Big data tools must be applied to the scientific workflow to make
these data sets manageable and useful.  This includes all aspects of
the data management cycle: ingest, metadata management, bulk data
movement and storage, and accessibility for processing and analysis.
Further, we desire solutions that can be run \I{interactively}: that
is, while the scientist is operating the X-ray equipment, data processing
operations proceed on available clusters and high-performance
computing (HPC) resources, so that experiment time is used optimally.
Thus, any errors in the experimental setup may be quickly detected and
corrected, and interesting features may be identified.  Today,
data analysis commonly is performed weeks or months after
data is collected; we have demonstrated the ability to accelerate the
scientific cycle to minutes.

The first stage of X-ray scattering analysis is essentially an image processing
problem. X-ray detectors produce large numbers of images in standard
formats.  Scientific tools already exist for various types of analysis
but are constantly under development; they are typically designed for execution
on small-scale resources such as a laptop, and are often implemented in
high-level languages such as Matlab.  The analyses of individual images are
generally independent.  Thus, a \I{many-task} framework~\cite{Raicu2008a} that can
rapidly compose applications from such existing codes and run them at
high concurrency levels can address many computing problems in X-ray
image analysis.  As shown in our results, this approach attains high
performance without the need for additional low-level coding (such as MPI messaging).

We consider here all aspects of the X-ray science data
management cycle, but focus in particular on the problem of ingesting and organizing large data
sets for many-task processing on supercomputers such as the IBM Blue
Gene/Q (BG/Q).  These supercomputers can be used by a many-task framework
such as Swift~\cite{Swift_2011,SwiftT_2013}, but must provide high-efficiency access to data sets.
As our performance results show, we cannot achieve high performance if individual tasks
access the shared file system independently. An alternative, coordinated approach is required.

Our approach is to stage input data to compute-node-local filesystems,
such as the RAM disk on the BG/Q (or a solid-state disk on a
hypothetical system).  We load this data using a simple technique
built around MPI-IO shared file operations for maximal efficiency.
Once data staging is complete, scientific workflow tasks are
distributed to processors where they can perform local data
operations with high I/O rates.  Our approach thus combines a
\I{collective} phase for big I/O operations with a \I{loosely coupled}
phase consisting of independent data analysis tasks.  For interactive
analysis, the staged data could be reused over several
human-in-the-loop cycles (although we do not address that here).

\begin{figure}[ht]
  \begin{center}
    \includegraphics[scale=0.75]{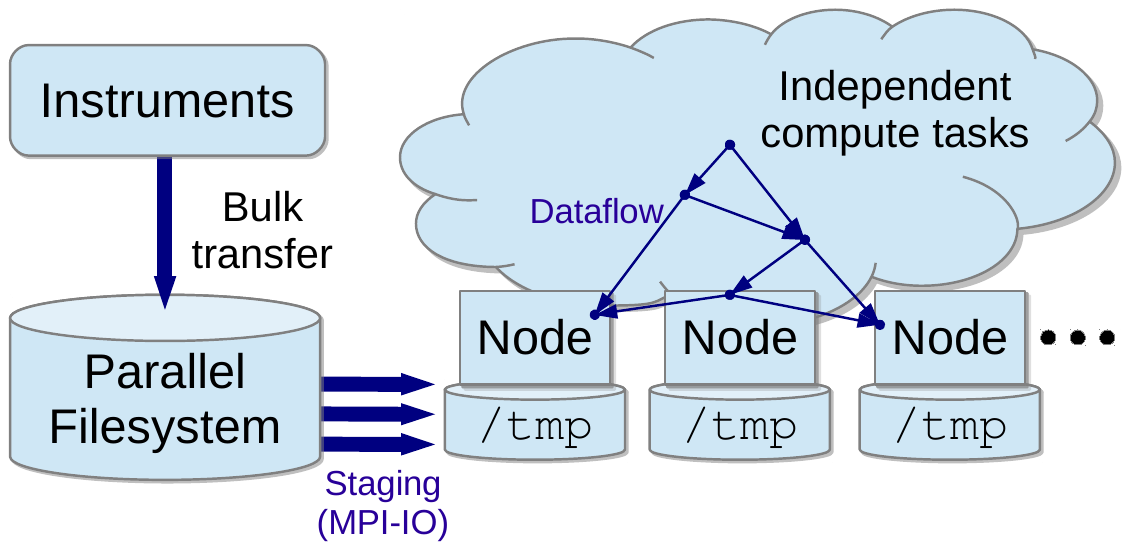}
    \caption{Overview of I/O staging; many-task workflow.
      \label{figure:overview}}
  \end{center}
\end{figure}

\FIG{overview} is a simplified diagram of our framework.  Scientific
instruments, such as X-ray detectors, produce large quantities of data
that are streamed to the parallel filesystem (e.g., GPFS) of the HPC
machine.  When an analysis run is triggered, this data is staged to
RAM disks (e.g., \T{/tmp}) on each compute node (not each core).  This
operation uses available collective I/O libraries (e.g., MPI-IO
\T{MPI\_File\_read\_all()}). Then, the workflow proceeds, with
independent tasks executing to process local data.

This approach offers many benefits.  First, the underlying
\I{scientific codes are not modified}; the code is not changed or even linked with a new library.
They simply operate on data in the local directory instead
of a shared filesystem directory.  Second, we provide a high-level
interface to the data staging operation---the user does not use
MPI-IO directly.  Third, the user workflow benefits from the use of a full-featured
language, capable of expressing MapReduce and more general patterns,
including loops, conditionals, and functions.

The remainder of this paper is organized as follows.  In\SL{hedm}, we
present an overview of a feature application in this paper:
high-energy diffraction microscopy.  In\SL{swift}, for completeness,
we present an overview of the Swift language, used to program the
workflows described here.  In\SL{hook}, we describe the new Swift I/O hook
feature that we introduce for I/O enhancement.  In\SL{workflow}, we
describe the HEDM workflows in more detail.  In\SL{performance}, we
provide performance results, and in\SL{summary}, we summarize our
results.

\section{Scientific overview: High Energy Diffraction Microscopy}
\label{section:hedm}

We review here the scientific application
that motivates our work in I/O optimization for many-task computing.
We note that the physical parameters of the scientific experiment have
a direct impact on various computing requirements (data size, available
concurrency, task granularity, etc.).

High-energy diffraction microscopy (HEDM)
\cite{HEDM_2011,HEDM_2012a,HEDM_2012b} is an important method for
determining the grain structure of metals.  It is performed at
specialized light sources such as the APS.  In the typical scientific
workflow, the scientist applies for a week of beam time and spends
that time gathering data.  Over the next several months, the data is
processed by using custom-built tools. In current practice,
HPC is rarely if ever applied during the
week of beam time.

Our work is intended to allow the use of HPC to analyze data quickly:
as it is produced by the experiment detectors or immediately after.
This integration of HPC into the scientific workflow has many potential
benefits. As we show below, it can permit rapid detection of and
recovery from errors.  It may provide feedback to the scientist during the run, to
improve the quality of results and the utility of precious beam time.
The overall result is to speed the process of scientific discovery,
even if the HPC is applied after beam time.

We focus here on the application of HPC to an HEDM application,
in an environment that encompasses an APS beam line, a small compute
cluster, and the Argonne BG/Q HPC system.  We use up to 64K cores of the BG/Q to provide near-real-time
feedback to APS beam users.

HEDM is a diffraction-based imaging technique that can non-destructively
determine the grain-level properties of polycrystalline materials. It yields
unique in situ 3D information which has previously only been available
through destructive, but more widely-availabe microscopy techniques,
such as
electron backscattering diffraction (EBSD). Hence, it is a valuable technique
for analyzing grain defects in advanced alloy materials, such as
those used in turbine blades for both energy (e.g., wind turbines) and
engine applications (e.g., jet engine turbines). The technique allows a
material sample, or even a manufactured part, to be studied in situ at
an APS beamline across a range of applied thermal and mechanical loading
conditions.  A polycrystalline material sample (typically a metal alloy)
is positioned within a high-energy ($E > 50\ \textrm{keV}$) X-ray beam, producing
forward-scattered X-rays that are collected by a range of detectors to
yield 2D material information.  A series of diffraction patterns is then
collected as the sample rotated about a single axis perpendicular to the
incident X-ray beam.  The diffraction patterns are analyzed by software
tools to reconstruct the 3D structure of the material, in order to determine the granular structure of material defects that can cause
component failure in fabricated parts.

Depending on the detector type and its placement with respect to the
sample, HEDM has two variants.  In the \I{near-field} (NF) variant, a
high-resolution detector ($\sim$1.5 $\mu$m pixel size) is places in
close proximity to the sample ($\sim$10 mm). In the \I{far-field}
(FF) variant, a medium resolution detector
($\sim$200 $\mu$m pixel size) is placed at a relatively larger
distance from the sample (up to 1 m).

The near-field HEDM technique \cite{HEDM_2006} works as follows. A line-focused X-ray
beam and a detector are used to collect diffraction data from a 2D cross-section
(``layer'') of a rotating polycrystalline sample. 2D
TIFF images, each 8 MB in size, are collected at each
angle of rotation, typically 360 to 1,440 angles per layer, and for
multiple detector distances. The computational analysis is split into
two stages. In Stage 1, the diffraction images are binarized to
detect pixels with diffraction signal. In Stage 2, a grid is simulated
on the virtual 2D cross-section of the sample, and diffraction signals
at each point on the grid are calculated and compared with the
diffraction images. This computation is parallelized at the grid point level
for a total of $\sim$10{$^5$} points per layer,
enabling the concurrent use of tens of thousands of processor cores.

\begin{figure}[ht]
  \begin{center}
    \includegraphics[scale=0.5]{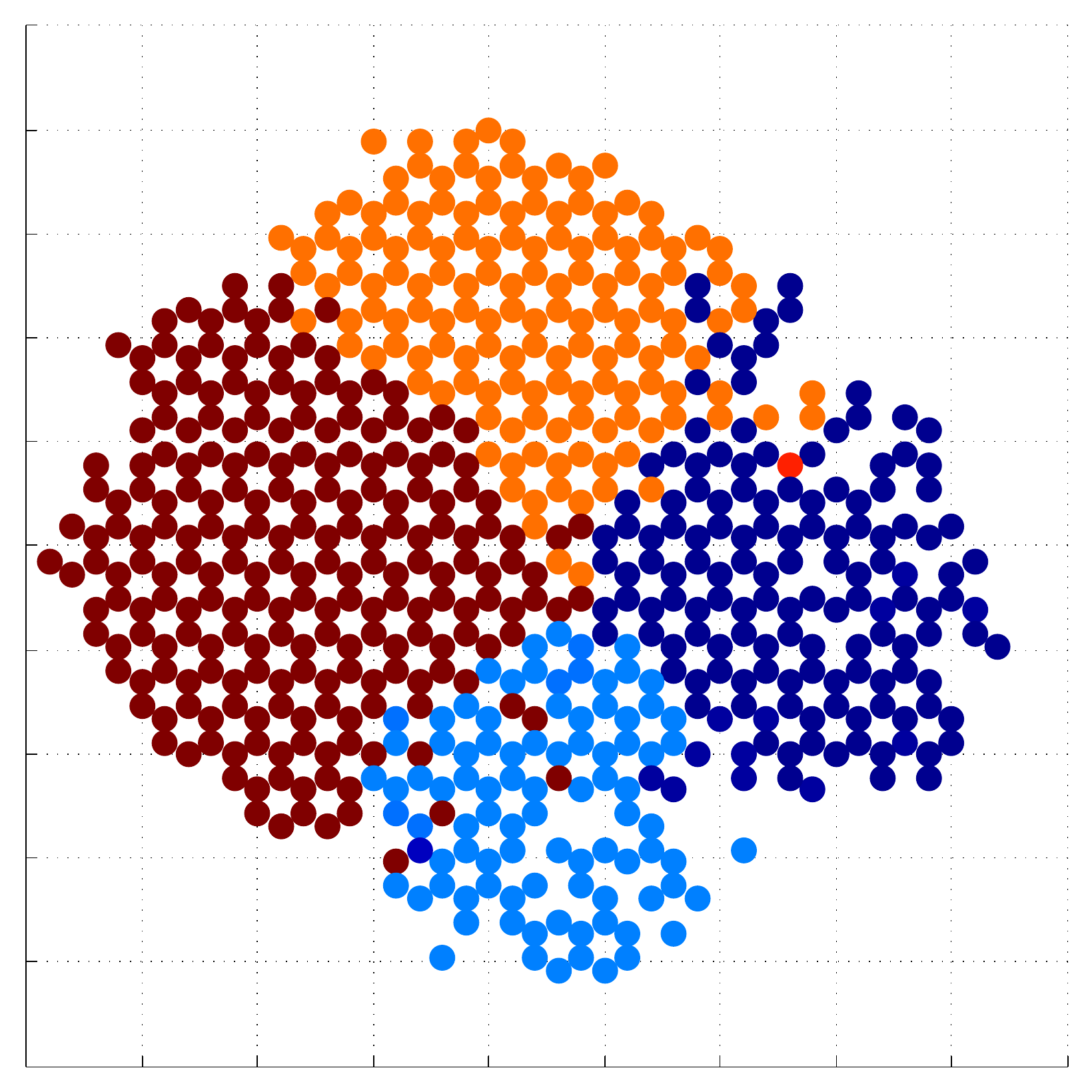}
    \caption{NF-HEDM grain identification for cross section of gold
      wire.  Each grid square is 5~$\mu$m.  Sample points of
      the same color have the same crystallographic orientation and
      are thus the same grain.
      \label{figure:gold-grains}}
  \end{center}
\end{figure}

The NF-HEDM sample in \FIG{gold-grains} shows the
cross section of a roughly round gold wire.  Each point in the
hexagonal grid is displayed as a colored dot (the grid is a hexagonal
prism in 3D).  The four colors correspond to the four distinct grains
identified in this sample, providing rich information about grain
characteristics.  This small example grid has 601 points, corresponding to that
many tasks; each task runs for about 10 minutes.

\begin{figure}[ht]
  \begin{center}
    \includegraphics[scale=0.5]{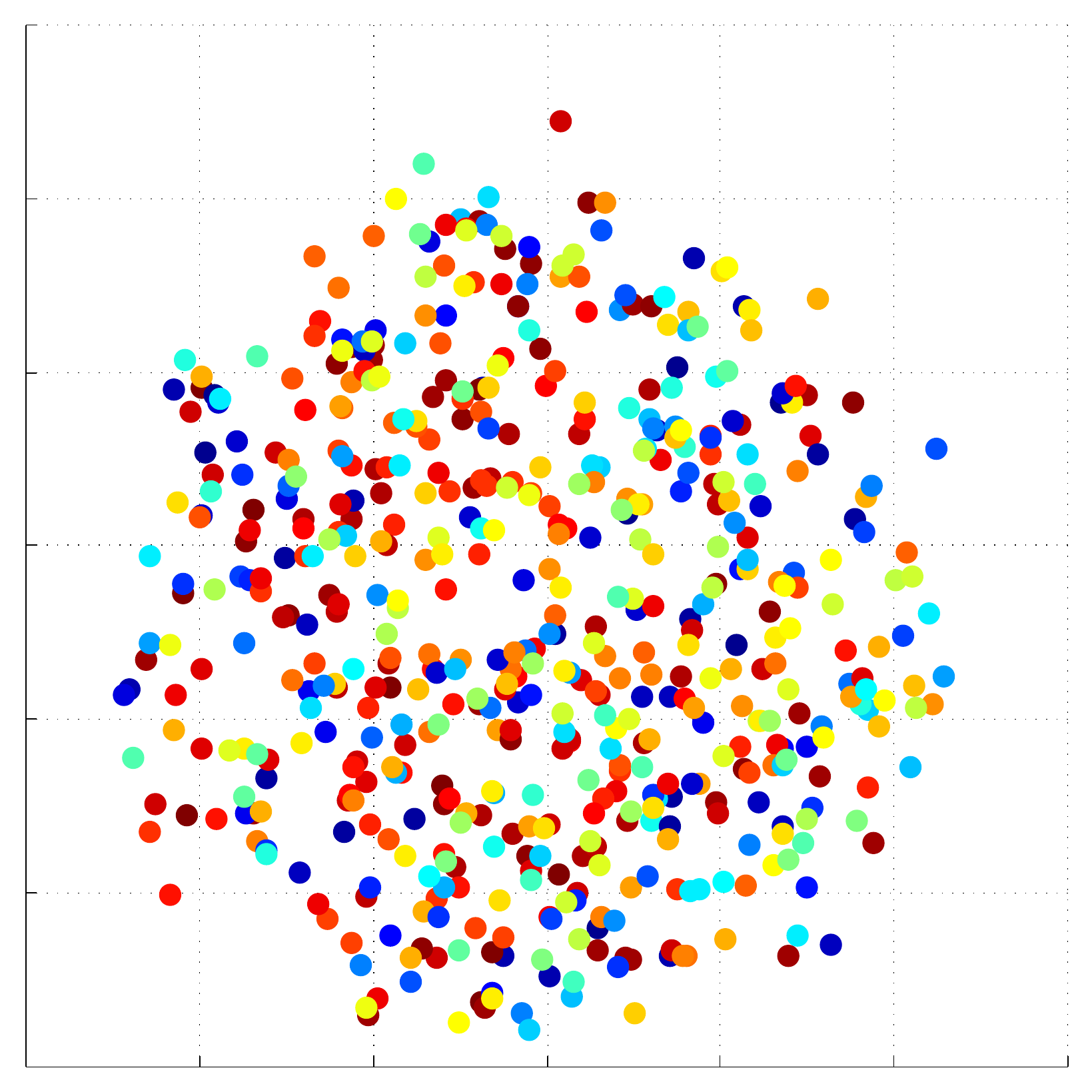}
    \caption{FF-HEDM grain identification for cross section of
      experimental material.  Each grid square is 200$\mu$m.  In this
      technique, only grain centers are identified, and grain sizes
      vary widely, producing this apparently unstructured image.
      \label{figure:ff-grains}}
  \end{center}
\end{figure}

In far-field HEDM, a ``box'' beam with a rectangular cross-section
illuminates a \I{volume} in the polycrystalline sample. The sample is rotated
to capture multiple diffraction spots from each grain (contiguous region
with the same crystal orientation) in the sample. A diffraction image,
8 MB in size, is recorded for 360 to 1,440 positions during the sample
rotation. In the first step of analysis, the diffraction images are
segmented, and properties of the diffraction spots are calculated. In the
second step, the diffraction spots are assigned (called ``indexing'')
as belonging to grains, and properties of the grains are calculated.
By imaging a volume instead of a cross section, FF-HEDM can image more
material per unit time, but at a lower information level than
NF-HEDM.

An FF-HEDM sample is shown in \FIG{ff-grains}.  This shows the
cross-section of a roughly round wire of an experimental material.  In
contrast to the NF-HEDM diagram, each dot here represents a grain
center; only one point is shown per grain.  This level of information
is sufficient to measure material response to stress and deformation
and provide feedback to material manufacturing techniques (e.g.,
annealing).  This diagram has 572 grid points.  The 4,109 tasks for
this case are described further in\SL{ff-hedm-stage-2}.

\section{Overview of the Swift language}
\label{section:swift}

Swift~\cite{Swift_2011} is a dataflow language for scientific
computing, commonly used in cluster, cloud, and grid environments.
Its recent re-implementation, Swift/T~\cite{SwiftT_2013}, is designed
for use on HPC systems. It operates by translating the Swift program
into a format for execution on an MPI-based
runtime~\cite{Swift_MPI_2013}, providing performance of up to 1.5
billion tasks/s on 512K cores of a Cray XE6 system.  This performance
is accomplished with a combination of compiler
optimizations~\cite{STC_2014}, an enhanced workflow enactment
technology called Turbine~\cite{Turbine_2012}, and a high-performance
load balancer called ADLB~\cite{ADLB_2010}.

Swift is an \I{implicitly parallel} language---all expressions may be
evaluated concurrently, limited only by dataflow ordering and
available processor.  This allows for a natural parallel programming
style, while still allowing for conventional constructs such as
\T{for} loops and \T{if} conditionals, as well as other features that
allow good use of practical machines~\cite{Swift_2014}.  Users may
link to user code in compiled (C, C++) or scripting languages (e.g.,
Python, Julia) in \I{leaf functions}, external code that consumes and
produces Swift data (numbers, strings, byte arrays, etc.).  Data is
moved through the Swift workflow over MPI but does not require the
user to write MPI code.

As a language, Swift has many features that promote big data
programming.  First, Swift has a rich feature set for
typical filesystem tasks: it supports simple calls to common shell
programs, provides glob and other typical operations, and includes a
\I{mapper} concept to \I{map} Swift variables to filesystem objects
(or URLs).  Second, Swift provides locality and work type features to
\I{send work to data} and reduce data movement.  Third, the dataflow
programming model can express a wide range of data analytics
algorithms elegantly and make them \I{extensible} through the addition of
Swift code.

For example, consider the popular MapReduce~\cite{MapReduce_2004}
framework. It would be interesting to extend this framework to support loops, conditionals, or
subworkflows, but such extensions would require significant changes to the whole MapReduce
implementation.  A Swift implementation of MapReduce allows such
changes to be made and tested piecemeal by adding lines to the script.
We show in \FIG{mapreduce-code} a
simplified Swift implementation of MapReduce, with a single merge
bucket. (A complete MapReduce implementation is the subject of
  work under preparation for submission elsewhere.)
 In this code, \T{find\_file()}, \T{map\_function()}, and
\T{merge\_pair()} are user-written leaf functions that consume and
produce ordinary files. These functions can be implemented in any language
and presented to Swift.

\begin{figure}[t]
  \begin{center}
    
\begin{center}
\scriptsize

\begin{tabular}{r|l}
 1 & {\tt \textbf{\textcolor{swiftbuiltincolor}{main}}\ \{} \\
 2 & {\tt \ \ \textcolor{swiftbuiltincolor}{file}\ d[];} \\
 3 & {\tt \ \ \textcolor{swiftbuiltincolor}{int}\ N\ =\ string2int(argv(\textcolor{swiftstringcolor}{"N"}));} \\
 4 & {\tt \ \ \textcolor{swiftcommentcolor}{//\ \textrm{\ Map\ phase}}} \\
 5 & {\tt \ \ \textbf{\textcolor{swiftbuiltincolor}{foreach}}\ i\textbf{\textcolor{swiftbuiltincolor}{\ in}}\ [0:N-1]\ \{} \\
 6 & {\tt \ \ \ \ \textcolor{swiftbuiltincolor}{file}\ a\ =\ find\_file(i);} \\
 7 & {\tt \ \ \ \ d[i]\ =\ map\_function(a);      } \\
 8 & {\tt \ \ \}                                  } \\
 9 & {\tt \ \ \textcolor{swiftcommentcolor}{//\ \textrm{\ Reduce\ phase}}} \\
10 & {\tt \ \ \textcolor{swiftbuiltincolor}{file}\ final\ {\textless}\textcolor{swiftstringcolor}{"final.data"}{\textgreater}\ =\ merge(d,\ 0,\ tasks-1);} \\
11 & {\tt \}                                      } \\
12 & {\tt                                         } \\
13 & {\tt (\textcolor{swiftbuiltincolor}{file}\ o)\ \textbf{merge}(\textcolor{swiftbuiltincolor}{file}\ d[],\ \textcolor{swiftbuiltincolor}{int}\ start,\ \textcolor{swiftbuiltincolor}{int}\ stop)\ \{} \\
14 & {\tt \ \ \textbf{\textcolor{swiftbuiltincolor}{if}}\ (stop-start\ ==\ 1)\ \{} \\
15 & {\tt \ \ \ \ \textcolor{swiftcommentcolor}{//\ \textrm{\ Base\ case:\ merge\ pair}}} \\
16 & {\tt \ \ \ \ o\ =\ merge\_pair(d[start],\ d[stop]);} \\
17 & {\tt \ \ \}\ \textbf{\textcolor{swiftbuiltincolor}{else}}\ \{} \\
18 & {\tt \ \ \ \ \textcolor{swiftcommentcolor}{//\ \textrm{\ Merge\ pair\ of\ recursive\ calls}}} \\
19 & {\tt \ \ \ \ n\ =\ stop-start;               } \\
20 & {\tt \ \ \ \ s\ =\ n\ \%\ 2;                 } \\
21 & {\tt \ \ \ \ o\ =\ merge\_pair(merge(d,\ start,\ \ \ \ \ start+s),} \\
22 & {\tt \ \ \ \ \ \ \ \ \ \ \ \ \ \ \ \ \ \ \ merge(d,\ start+s+1,\ stop));} \\
23 & {\tt \ \ \}                                  } \\
24 & {\tt \}                                      } \\

\end{tabular}
\end{center}

    \caption{MapReduce-like application expressed in Swift.
      \label{figure:mapreduce-code}}
  \end{center}
\end{figure}

\begin{figure}[t]
  \begin{center}
    \includegraphics[trim=0.1in 0 0 0,clip,scale=0.8]{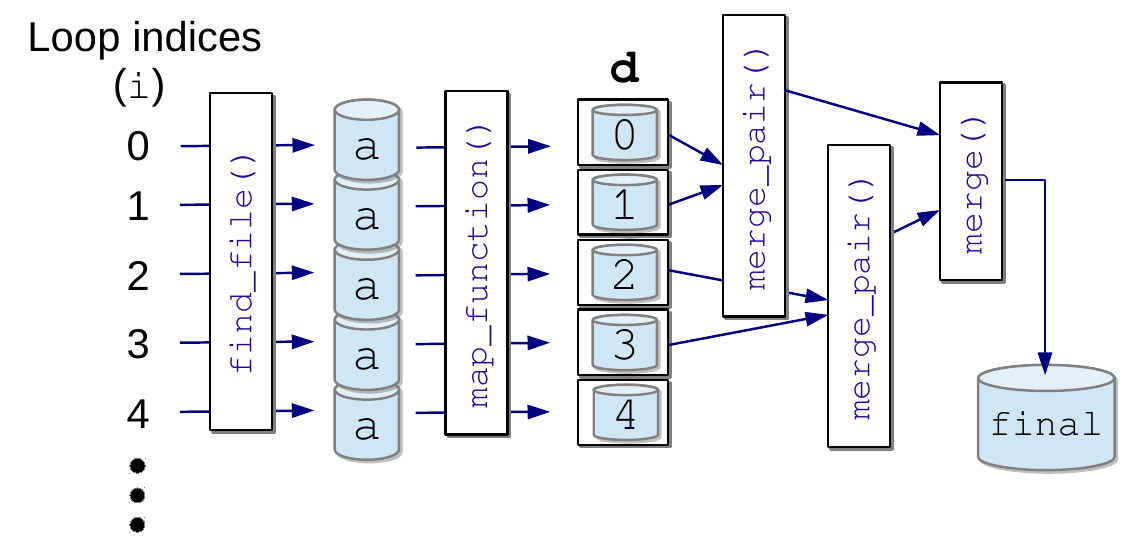}
    \caption{Dataflow diagram of MapReduce-like application.
      \label{figure:mapreduce-diagram}}
  \end{center}
\end{figure}

A dataflow diagram of the MapReduce implementation is shown in
\FIG{mapreduce-diagram}. The application proceeds as follows. First,
an integer \T{N} is obtained from the user command line (line 3).  The
\I{map phase} operates simply: for each value of \T{i} from 0 to
\T{N-1}, a file \T{a} is obtained and processed with the
\T{map\_function()} (lines 6--7).  The result is stored in array \T{d}.
The map functions operate concurrently and are automatically load
balanced, limited only by available processors.  The \I{reduce phase}
(line 10) begins as soon as mergeable data is available.  This is
implemented here as a recursive (line 21) pairwise reduction on the
contents of \T{d}.  \T{merge\_pair()} is called on consecutive entries
in \T{d} (line~16), then recursively reduced pairwise up to the top call to
Swift function \T{merge()}.  The resulting file (variable \T{final}) is
stored in physical file \T{final.data} (line 10).

Note that this dataflow expression of simplified MapReduce does not
have a barrier between the map and reduce phases
(see~\cite{MR_Barrier_2013}).

\section{The Swift I/O hook}
\label{section:hook}

As described in\SL{swift}, Swift may be used to
express complex data analytics workflows, make use of data locations,
and operate with high performance.  In typical HPC
settings, however, data is stored in a shared parallel filesystem and is not
resident on compute nodes (as in HDFS).  Thus, data must be quickly
\I{staged} to compute nodes for processing.  Then, Swift execution can
proceed, operating on node-local data as well as shared data (if
desired).

This feature can be used to address filesystem congestion problems due
to bandwidth \I{or} many-small-file congestion.  For example, on an HPC
system such as the BG/Q, plain executables are already broadcast efficiently
to all compute nodes prior to execution by the standard BG/Q operating
system.  Swift scripts, however, may
desire to make use of a large collection of small Python scripts or
C++ shared libraries.  The operating system cannot automatically load
these on compute nodes efficiently.  The Swift I/O hook can be used to
pre-stage any such data: scripts, shared objects, configuration files,
and so on, thus improving application start time and reducing impact on
other system users.  Bandwidth saving is a simpler benefit; we
eliminate unnecessary duplicate bandwidth consumption.  The Swift I/O
hook is thus designed to reduce I/O traffic for both small and large
file operations, including its operations used in its implementation.

The Swift I/O hook is implemented in the following way.  First, a
\I{leader communicator} is automatically constructed by the runtime.
This MPI communicator consists of exactly one ADLB worker process per
node.  If a \I{leader hook} (e.g., the I/O hook) is provided in an
environment variable, it is executed by each process in the the leader
communicator.  More complex scripts can be distributed by using the
leader hook environment variable and then entered by the hook script.

We show in \CODE{hook} the code for an example Swift I/O hook.  This fragment,
which is evaluated by the interpreter in the Swift runtime, defines a list of
broadcast definitions, each of which targets
(\T{broadcast~to}) a node-local directory location.  The file list can
contain glob patterns.

\begin{figure}[h]
  
\begin{center}
\scriptsize

\begin{tabular}{r|l}
 1 & {\tt \textbf{broadcast\ to}\ /tmp\ \textbf{files}\ \{} \\
 2 & {\tt \ \ $\sim$/dataset-1/*.cfg              } \\
 3 & {\tt \}                                      } \\
 4 & {\tt                                         } \\
 5 & {\tt \textbf{broadcast\ to}\ /tmp/bulk\ \textbf{files}\ \{} \\
 6 & {\tt \ \ \ \ $\sim$/dataset-1/bulk/file1.index} \\
 7 & {\tt \ \ \ \ $\sim$/dataset-1/bulk/file2.index} \\
 8 & {\tt \ \ \ \ $\sim$/dataset-1/bulk/*.bin     } \\
 9 & {\tt \}                                      } \\

\end{tabular}
\end{center}

  \caption{Example Swift I/O hook specification.
    \label{code:hook}}
\end{figure}

In order to avoid filesystem metadata contention, the file list is \I{also}
broadcast by using \T{MPI\_Bcast()}, with the result that only one process performs any
globs. (A naive implementation would simply run the glob on each
process to obtain the list of transfers to perform, congesting the
shared filesystem.)

The hook is then used from the command line as follows:

\begin{center}
\scriptsize

\begin{tabular}{r|l}
 1 & {\tt SWIFT\_IO\_HOOK=\$(cat\ hook)\ \textbf{swift-t}\ options\ program\ ...} \\

\end{tabular}
\end{center}

At execution time, the Swift/T implementation processes the I/O hook
shortly after initializing its communicators.  Swift/T performs the
globs on one rank, broadcasts the list of files to transfer, then uses
\T{MPI\_File\_read\_all()} to make read-only replicas of each
file on each node-local file system.

\paraheader{Future directions}
The leader hook is a generic mechanism
that may be generalized for more complex functionality in
future work.  The leader communicator, a new feature, is derived from the
Swift/T \I{hostmap} functionality, which maps host names (nodes) to
MPI ranks, allowing for location-specific programming at the workflow
level~\cite{Swift_2014}.

The leader hook is actually a Tcl fragment and thus in essence provides an
extension language for Swift/T.  In principle, the user can use this language to
access the leader communicator and program arbitrary operations.
Since this is an error-prone process, however, we provide the
high-level wrapper syntax shown in~\CODE{hook} (which is still Tcl
code).

\section{Application description}
\label{section:workflow}

Having discussed the scientific application of interest in\SL{hedm}
and our software system in\SL{swift} and\SL{hook}, we now describe
the motivating scientific workflow and practical details that motivate
our solution.

\subsection{The scattering science workflow}

Scattering science is a complex process that involves instruments,
computers, and humans in the loop.  The APS is a major scientific
investment, hosting 5,000 users per year~\cite{APS_WWW}.  Users
typically visit the laboratory for one week, during which they have
access to beam resources 24 hours a day.  Typically, data is collected
on portable hard drives that are carried to the user's home
institution for processing and analysis.  High-resolution detectors
can produce 15 TB per week, but typical users fill less than one hard
drive.

Our high-level goal is to improve this human-in-the-loop workflow by
delivering enough computing power to the application so that analysis
can be performed \I{during beam time}, that is, while the visitor is
present at Argonne.  The HEDM application considered in this paper could
use $\sim$22 M CPU core hours per week, a quantity that can be
obtained only on extreme-scale supercomputers.  However, the process is
additionally constrained by the size of data to be moved to the
supercomputer and the amount of data to be transferred to compute
nodes---these are the big data problems.  In the most recent run,
considered here, 2 TB of data were collected in two days.  Without our
optimizations, roughly half of that time would have been spent in I/O and half in
computing.

\begin{figure}
  \begin{center}
    \includegraphics[scale=0.8]
                    {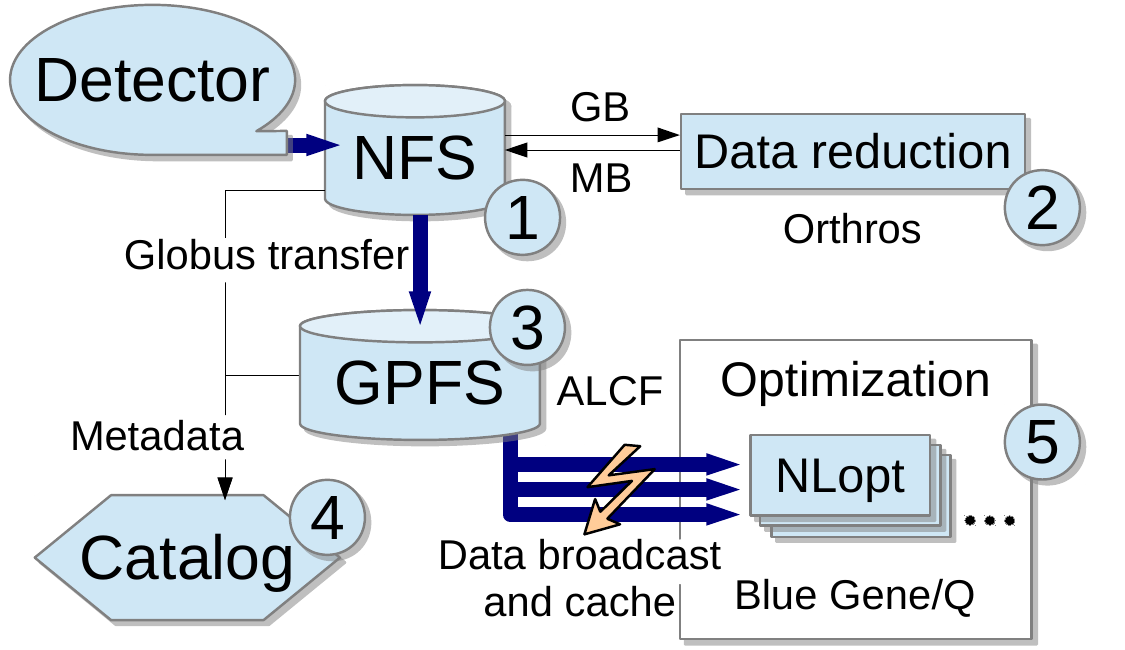}
    \caption{Cross-lab APS to ALCF workflow.
      \label{figure:nfhedm-workflow}}
  \end{center}
\end{figure}

Additionally, the human-in-the-loop nature of the runs motivates I/O
advancements.  It is desirable for the investigator to be able to
visualize detector data and detect anomalies or make decisions as soon
as detector data is available.  In order to keep up with the detector data
generation rate, the entire workflow must complete in five minutes.  Thus, I/O overheads cannot be
amortized into long running jobs, but must operate at near-interactive
rates.  Our
methods have been able to reduce run times for experimental datasets
to below that rate, given enough compute cores.  (Reservations on
BG/Q systems have been allocated to users on the beamline, allowing
interactive use without queues.)

\subsection{Workflow practicalities}

The raw TIFF files are reduced to binary files containing only
information about the diffraction signal of the detector. Because of the
sparse nature of the data, each 8 MB raw file can be reduced to an
$\sim$1 MB binary file. This reduction is performed by using 320 cores
on Orthros, each using about 2 min of CPU core time. These binary
files are fed into an orientation detection program to map the
orientation of each point in a grid in the 2D cross-section of the
sample. Two approaches are possible for this step.  Using 320 cores on
Orthros, at about 30 s for each grid point (a total of 100K grid
points), we can obtain results in three hours.  Using 10,000 cores on
the Mira BG/Q or similar parallel resource, we can scale this
workflow to obtain results in less than five minutes.

The $\sim$10 MB output file contains information about the
orientation of each point in a grid in the 2D cross-section of the
sample.  All analysis software has been developed in-house in Sector
1, implemented in C. Parallel processing of the analysis routines is
implemented by using the Swift parallel scripting language.

Figure~\ref{figure:nfhedm-workflow} shows the
NF-HEDM workflow in detail.  The detector produces data on an NFS installation
at the APS~\circled{1} and large numbers of data reduction jobs are run
on the local cluster, Orthros~\circled{2}.  The resulting data is moved
via the Globus transfer service~\cite{FosterGO2011} to Argonne Leadership Computing Facility (ALCF)
storage~\circled{3}, and recorded in a metadata
catalog~\circled{4}~\cite{Catalog_2014}.  Finally, the HPC component
runs~\circled{5}, in which a large batch of hundreds of thousands of
optimization operations are performed rapidly across tens of thousands
of CPU cores of a BG/Q.

In this model, C analysis code developed for this work is linked to
the NLopt optimizer library and the GNU scientific library.  These C
code tasks are grouped into a large Swift script, which compiles into
an MPI program for execution on a large HPC resource.

\subsection{Code fragments}

As a concrete example, we show in \CODE{nfhedm-stage-2} the Swift code
for stage 2 of the NF-HEDM application.  This program takes as
arguments input parameter file and output microstructure file names,
as well as a range of grid points to analyze, allowing for
variable-sized runs.  Then, each grid point (row) is analyzed in
parallel with \T{FitOrientation()}, a C function that uses NLopt to
fit the grain.  These tasks are eligible to run concurrently with
automatic load balancing due to the Swift \T{foreach} loop.  The C
function requires all input data specified by the parameter file and
distributed by the Swift I/O hook; the output is inserted into the
microstructure file.

\begin{figure}
  \begin{center}
    
\begin{center}
\scriptsize

\begin{tabular}{r|l}
 1 & {\tt \textbf{\textcolor{swiftbuiltincolor}{main}}\ \{} \\
 2 & {\tt \ \ parameterFile\ =\ argv(\textcolor{swiftstringcolor}{"p"});} \\
 3 & {\tt \ \ microstructureFile\ =\ argv(\textcolor{swiftstringcolor}{"m"});} \\
 4 & {\tt \ \ start\ =\ toint(argp(1));           } \\
 5 & {\tt \ \ end\ \ \ =\ toint(argp(2));         } \\
 6 & {\tt \ \ \textbf{\textcolor{swiftbuiltincolor}{foreach}}\ row\textbf{\textcolor{swiftbuiltincolor}{\ in}}\ [start:end]\ \{} \\
 7 & {\tt \ \ \ \ \textbf{FitOrientation}(parameterFile,\ row,} \\
 8 & {\tt \ \ \ \ \ \ \ \ \ \ \ \ \ \ \ \ \ \ \ microstructureFile);} \\
 9 & {\tt \ \ \}                                  } \\
10 & {\tt \}                                      } \\

\end{tabular}
\end{center}

    \caption{Swift/T fragment for NF-HEDM stage 2.
      \label{code:nfhedm-stage-2}}
  \end{center}
\end{figure}

\section{Performance}
\label{section:performance}

We next present performance data for the HEDM
applications when running under our system.  Cluster timings were obtained on
\I{Orthros}, a 320-core x86 cluster at the APS. An Orthros node has 64 AMD
cores running at 2.2 GHz.  HPC results were
obtained on the BG/Q systems at the ALCF.  Each BG/Q node has 16 cores (64 threads) in the
PowerPC A2 architecture running at 1.6 GHz. Runs at or below 8,192
cores were performed on the ALCF's smaller BG/Q, \I{Cetus}; runs above that were performed on
the larger
\I{Mira} system.  Both BG/Q systems run GPFS~\cite{GPFS_2002}; the
installation supports a peak I/O performance of 240 GB/s~\cite{BGQ_IO_Glean_2014}.

\newcommand{\plotsize}{0.5}

\subsection{NF-HEDM: Data reduction step (cluster)}

The data reduction step involves, first of all, a median calculation on each pixel of the
detector, using all images. Then, independently on each image, the following computations
are performed: a median
filter, followed by a Laplacian of Gaussian filter to determine
the edges of the diffraction spots; a connected components
labeling step; and a flood fill operation to retrieve information regarding all
useful pixels in the image. When run on Orthros at our maximum allocation size of 320 cores,
this data reduction step required 106~s to process 736 images from two detector distances.

\subsection{NF-HEDM: Analysis step (HPC)}

The analysis step highlights the key contribution of this paper: the Swift I/O
hook.  We profile the use of the I/O hook as well as the overall
application.

\subsubsection{I/O profiling}

\begin{figure}
  \begin{center}
    \includegraphics[scale=0.9]{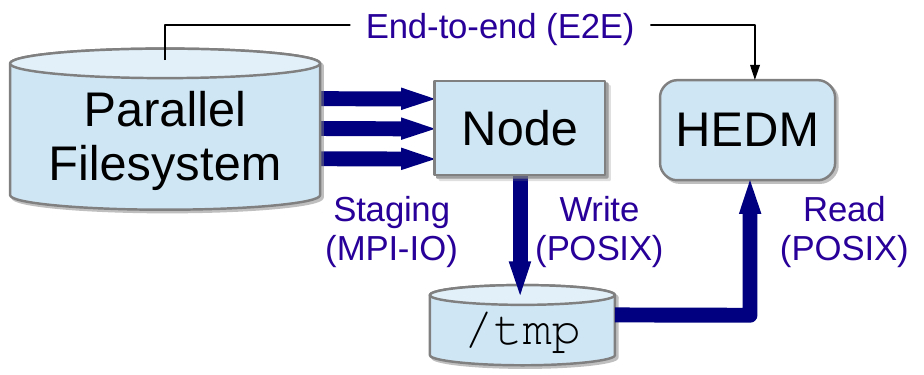}
    \caption{Definitions of data movement operations in the Swift I/O
      hook.  HEDM is the unmodified scientific application.
      \label{figure:io-model}}
  \end{center}
\end{figure}

We first isolate the input operations in order to study their performance
independently of the rest of the application. As shown in
\FIG{io-model}, we distinguish three I/O steps: \I{Staging}, \I{Write}, and \I{Read}.
The Swift I/O hook performs the \I{Staging} and \I{Write} steps, replicating data
from the filesystem to node-local storage. (Note that on the
  BG/Q the \T{/tmp} RAM disk is actually an I/O node service.)  The
\I{Read} phase is performed by the application task itself, by simply
reading from the appropriate directory specification (e.g., \T{/tmp}).

In our first experiment, we configured NF-HEDM to process a 577~MB
data set from GPFS. Each node requires a full replica of the data set
for use by tasks on that node. \PLOT{nf-stage} shows the performance
measured for the Swift I/O hook as it reads data from GPFS using
MPI/IO and writes the resulting data to node-local storage. At our
highest reported node count, 8,192 nodes, the system delivers data at
an aggregate rate of 134 GB/s.

\begin{figure}
  \begin{center}
    \includegraphics[scale=\plotsize]{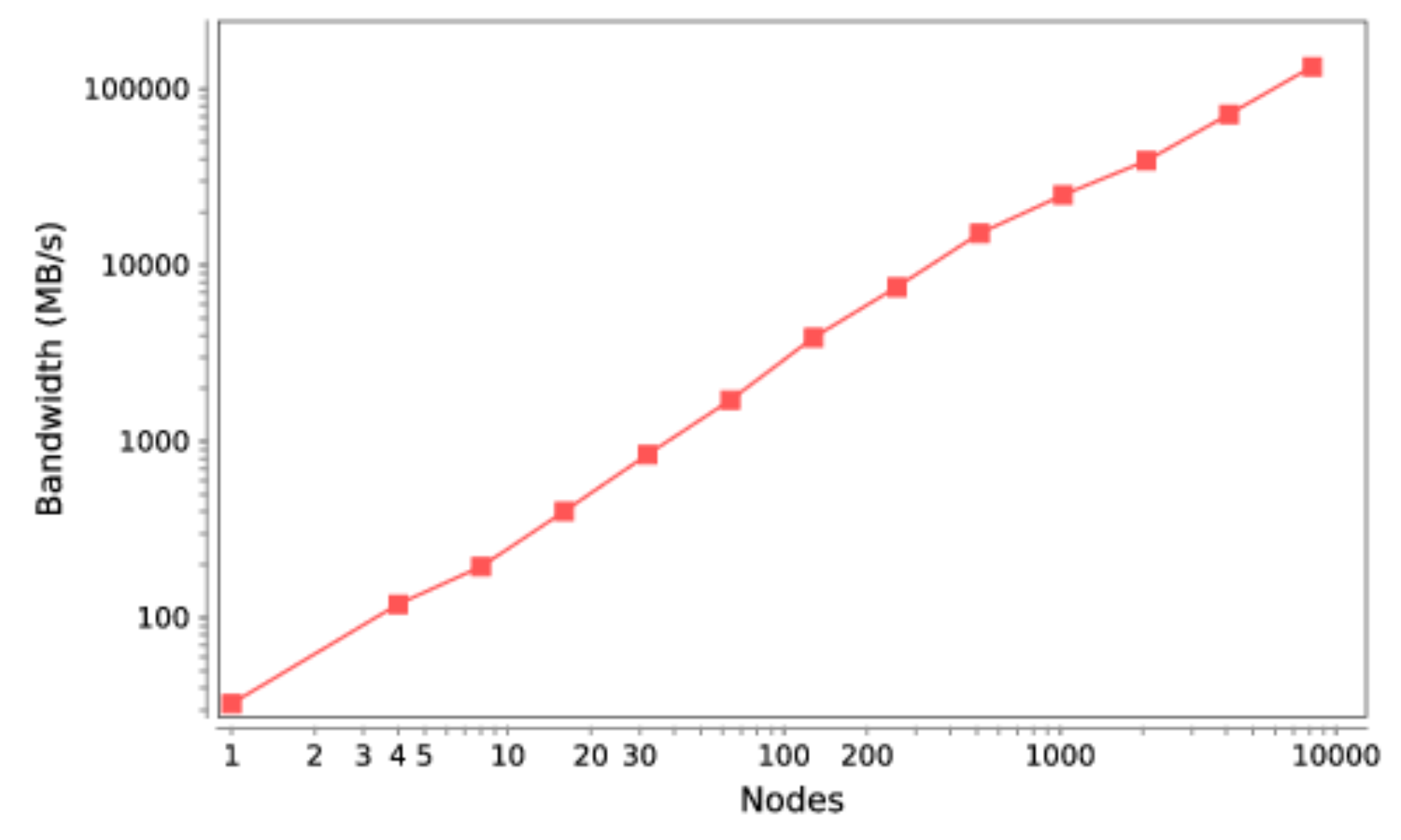}
    \caption{Staging+Write performance for NF-HEDM.
      \label{plot:nf-stage}}
  \end{center}
\end{figure}

The \I{Read} phase consistently takes 10.8~$\pm$~0.1~s regardless of
allocation size, for a per-process read bandwidth of 53.4
MB/s---comparable to a conventional node-local disk with ideal
scalability.

To determine the aggregate end-to-end input bandwidth achieved by our
approach, we add together the time taken for the joint \I{Staging}/\I{Write} and \I{Read}
phases.  The upper line in \PLOT{nf-input} shows the result, which reaches
101 GB/s on 8,192 nodes. In contrast, the original I/O approach, in which
each task reads input data independently from GPFS, without the use of collectives
achieves only 21 GB/s on 8,192 nodes.

\begin{figure}
  \begin{center}
    \includegraphics[scale=\plotsize]{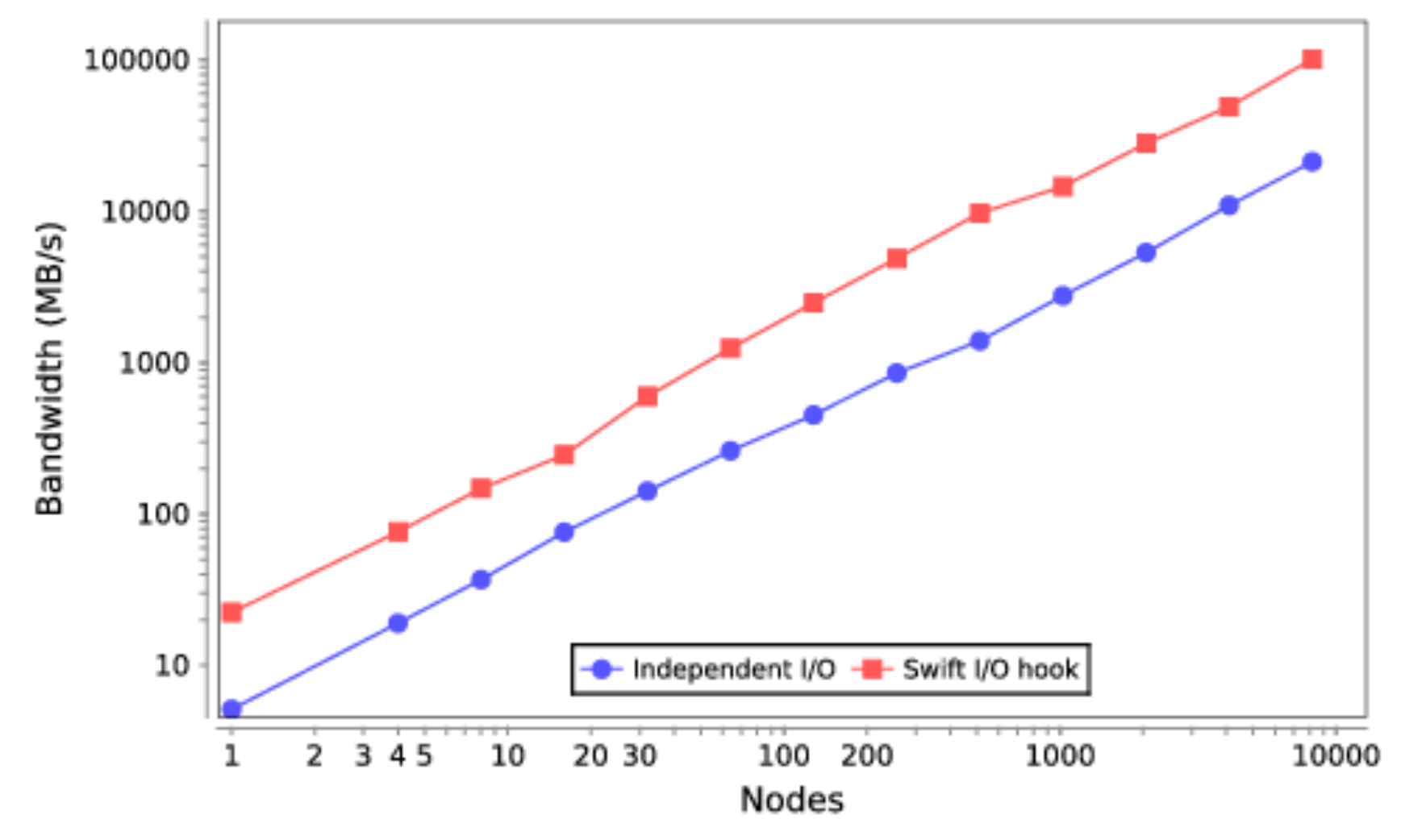}
    \caption{End-to-end performance for NF-HEDM.
      \label{plot:nf-input}}
  \end{center}
\end{figure}

We modified NF-HEDM to cache all inputs in application memory (for
each variable, tasks first check to see if it has already been read,
if not, they perform read operations to instantiate it).  Since
Swift/T reuses the same processes for subsequent tasks, HEDM tasks
after the first do not need to perform \I{Read} operations at all.  This
approach reduces input time to effectively zero for subsequent tasks.

In terms of wall time, the Swift I/O hook reduces the input time by a factor of 4.7, from
210~s to 46.75~s,
making data available to 8,192 nodes (524,288 hardware threads).

%
%
%
%

%
%

%
%

%
%
%
%
%
%

\subsection{FF-HEDM: Stage 1 (cluster)}

In FF-HEDM stage 1, each process loads a diffraction image (8~MB) and
characterizes all peaks in the image. The output is saved as a text file
($\sim$50 KB). We performed runs on 720 images, with each image being
processed in parallel. Depending on the number of diffraction spots in
each image, the processing time per image can vary from 5 s to 160 s. \PLOT{ff-1}
shows the scaling results for the 720 jobs on Orthros.

\begin{figure}
  \begin{center}
    \includegraphics[scale=\plotsize]
                    {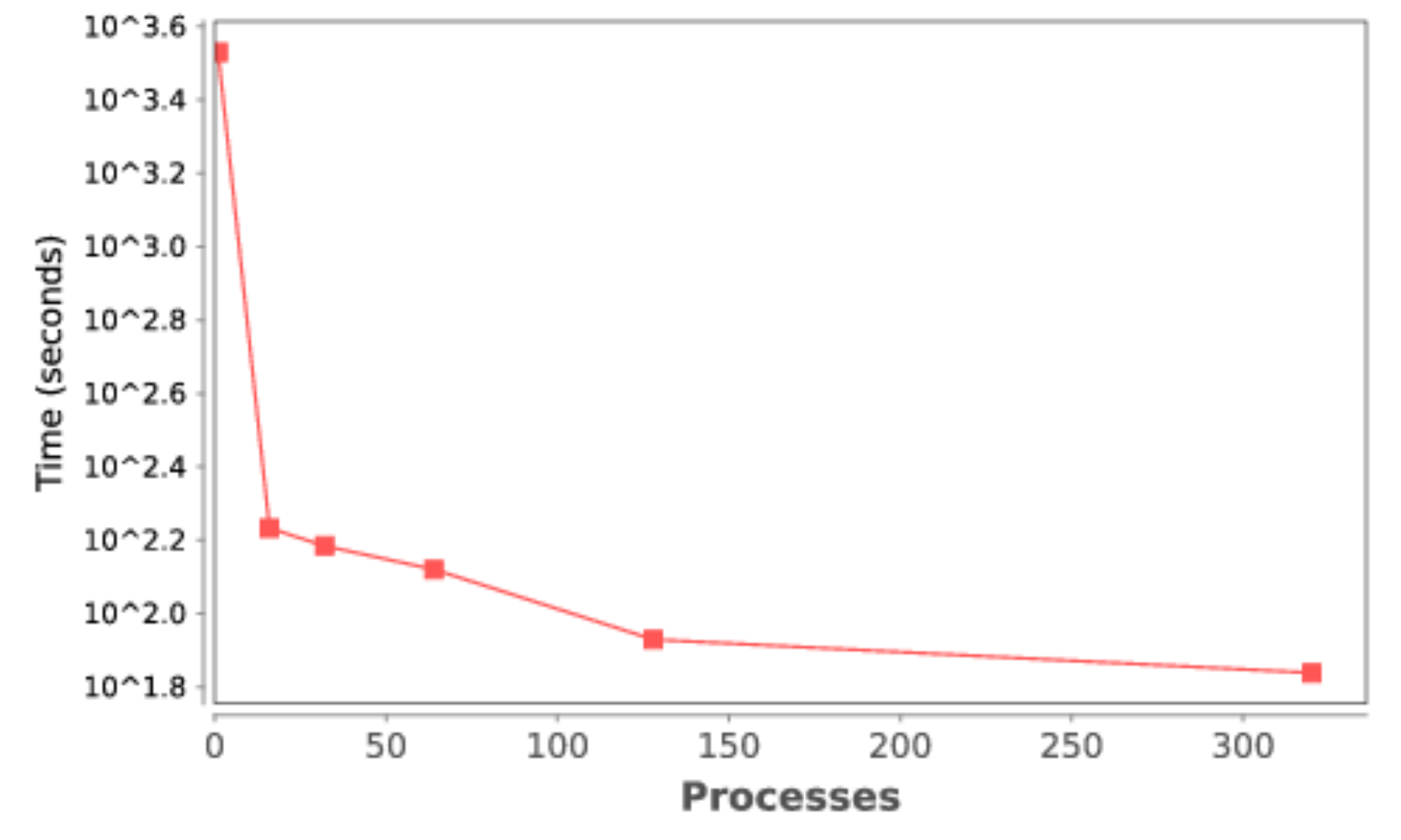}
    \caption{Makespan scaling result for FF-HEDM stage 1.
      \label{plot:ff-1}}
  \end{center}
\end{figure}

\subsection{FF-HEDM: Stage 2 (cluster)}
\label{section:ff-hedm-stage-2}

The number of tasks in this case is data-dependent, varying with the
number of grains within the sample volume illuminated by the diffraction
beam. For our experiments, we work with a sample that comprises 4,109 grains
and thus tasks, with the run-time
per task varying between 5 and 25 s, depending on the optimization landscape.
\PLOT{ff-2} shows the scaling results for the 4,109 jobs run on Orthros.

\begin{figure}
  \begin{center}
    \includegraphics[scale=\plotsize]
                    {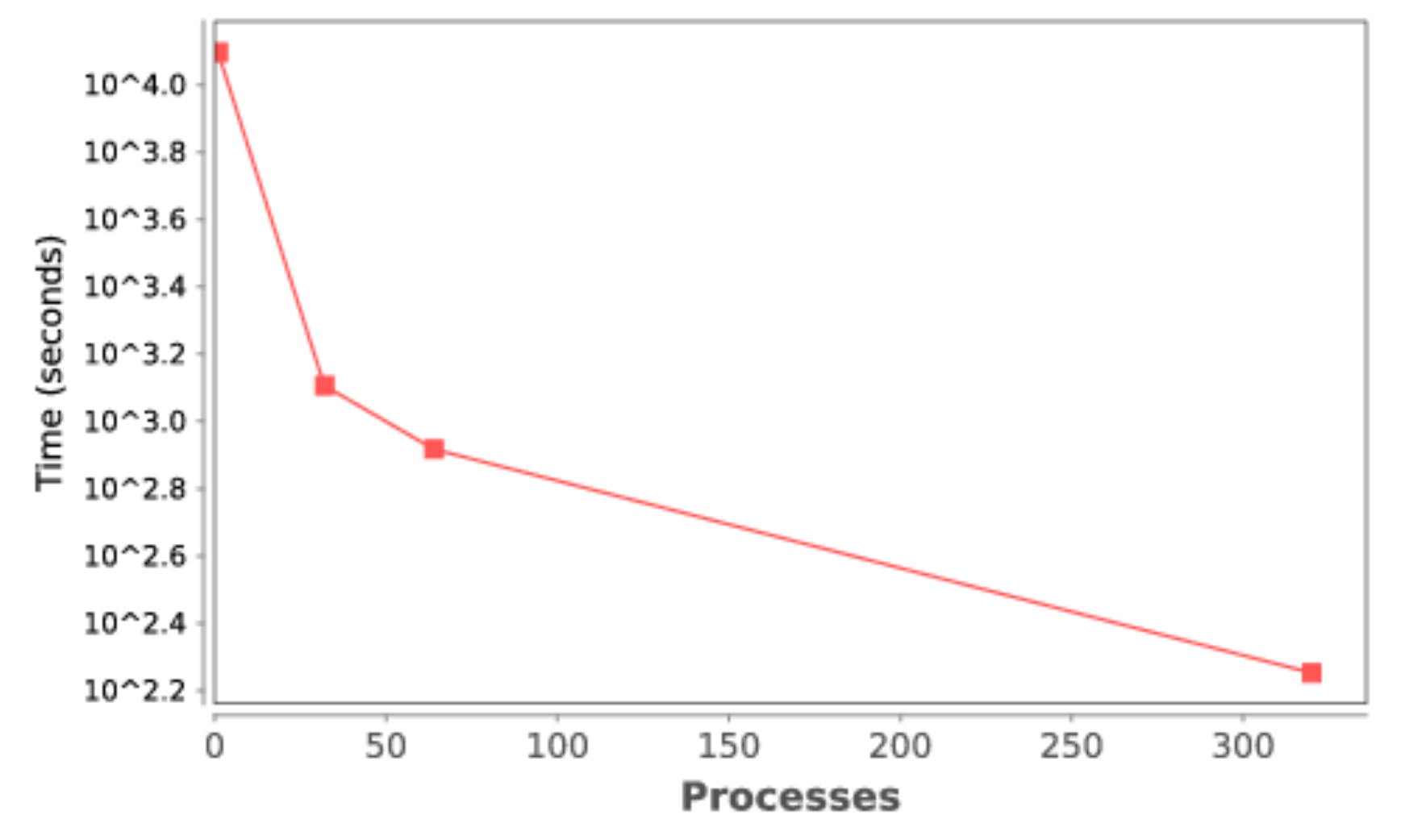}
    \caption{Makespan scaling result for FF-HEDM stage 2.
      \label{plot:ff-2}}
  \end{center}
\end{figure}

\section{Related work}
\label{section:related}

We aim in this work to make big data analysis feasible on HPC
systems such as the BG/Q. Our approach is to rapidly fill node-local storage with data
for in-place processing, and then launch analytics tasks using
a flexible, general-purpose dataflow programming model.  While this
approach is unique, much related work has been done.

The term scratch has multiple meanings but can be applied to the use of a
specialized, high-performance filesystem that trades reliability for
performance.  However, while scratch space may be intended to be used as cache, it is
not typically used as such for a variety of reasons, leaving users
with just another filesystem.  The Scratch as a Cache
system~\cite{ScratchCache_2009} acts on user-provided hints in the job
submission script, loading inputs into the fast filesystem
automatically, improving on workflow-oblivious least-recently used
(LRU) and other common techniques.  The interface to Scratch as a
Cache is similar to ours in that files are specified by the user
and moved to a high-performance store, improving on automated
techniques.  However, we focus on the use of node-local storage.

Workflow-aware storage/scheduling systems, including the Batch-Aware
Distributed Filesystem~(BAD-FS)~\cite{BADFS_2004} and the
Workflow-Aware File System (WASS)~\cite{WorkflowAwareFS_2012}), use
workflow information to make caching decisions, including an ad hoc
broadcast mechanism in WASS.  BAD-FS does not use collective I/O
operations.  WASS performance peaked at eight replicas (our results show
performance gains up to and beyond 8K replicas).  BAD-FS and
Freeloader~\cite{Freeloader_2006} use \I{cooperative caches} that
focus on write-once, read-many workloads (like ours) but
access \I{nearby} cached copies, instead of using
collectives. Other uses of node-local storage in HPC typically focus
on the caching and aggregation of write operations, rather than reads, as in our work.

Data Diffusion~\cite{DataDiffusion_2008} uses a distributed cache with
a distributed index to support many-task, data-intensive
Falkon~\cite{Falkon_2008} workloads for up to 128 processors; FusionFS
employs a similar architecture reporting larger results up to 1,024
nodes.  The Chirp~\cite{Chirp_2009} and AMFS~\cite{AMFS_2012} systems
feature broadcast operations in the cache storage system but use ad
hoc data distribution libraries and do not report performance results
at the scale of interest here.

Extract-Transform-Load (ETL) is a collection of methodologies and
technologies used in data warehouses to move data from
sources to processors.  It is responsible for ``(i) the extraction of
the appropriate data from the sources, (ii) their transportation to a
special-purpose area of the data warehouse where they will be
processed,'' and other operations~\cite{ETL_2009}.  ETL is
conventionally connected strongly to database applications and does
not emphasize collective operations or concurrency.  MapReduce has
been considered as an ETL system~\cite{MR_DB_2010} but not with a
highly concurrent preliminary staging operation.  MapReduce typically
operates on in situ (i.e., compute node resident) data, as does Dremel~\cite{Dremel_2010}.

MRAP~\cite{MRAP_2010} explored the idea of copying and converting data
from a parallel filesystem and scientific data format to a Hadoop
system in a MapReduce-friendly format. However, it does not consider
the use of collective I/O and focused on formatting issues.

\section{Summary}
\label{section:summary}

X-ray scattering is a broad and complex application area.  We have focused
here on a key part of the X-ray scattering analysis problem---moving data to processors
for computational analysis.  We described how the Swift language can
be used to express data analysis tasks in an elegant way, supporting
common data analytics patterns such as MapReduce as well as the scientific
pattern for HEDM.  We reviewed the scientific and
practical details of X-ray science and the HEDM application, and we provided
a detailed presentation of our key
contribution: the use of MPI-IO for data staging for big data
analytics.  We described the implementation
of this technique and presented performance results.
We showed that our HPC-oriented approach can be used to complete
the whole analysis phase of
the HEDM workflow extremely quickly ($\sim$5 minutes), which is
suitable for interactive scientific use.

\section*{Acknowledgments}

This material is based upon work supported by the U.S. DOE Office of
Science under contract DE-AC02-06CH11357 and by NSF award ACI 1148443.
Computing resources were provided in part by NIH through the
Computation Institute and the Biological Sciences Division of the
University of Chicago and Argonne National Laboratory, under grant S10
RR029030-01, and by NSF award ACI 1238993.  This research used
resources of the Advanced Photon Source and Argonne Leadership
Computing Facility at Argonne National Laboratory, which is supported
by the Office of Science of the U.S. Department of Energy under
contract DE-AC02-06CH11357.  We thank GE Global Research (Niskayuna,
USA) for financial support, technical discussions, and providing
industrial relevant materials for developing the HEDM techniques at
the Advanced Photon Source. We would like to thank R. M. Suter from
Carnegie Mellon University for providing the gold wire.  We thank Rob
Latham and Venkatram Vishwanath for consultation about I/O on the Blue
Gene/Q.

\bibliographystyle{abbrv}
\bibliography{swift}

\end{document}